**GENERAL ARTICLE**

# Artificial Intelligence, Values, and Alignment

**Iason Gabriel**[1]

**Abstract**
This paper looks at philosophical questions that arise in the context of AI alignment. It defends three propositions. First, normative and technical aspects of the AI alignment problem are interrelated, creating space for productive engagement between people working in both domains. Second, it is important to be clear about the goal of alignment. There are significant differences between AI that aligns with instructions, intentions, revealed preferences, ideal preferences, interests and values. A principle-based approach to AI alignment, which combines these elements in a systematic way, has considerable advantages in this context. Third, the central challenge for theorists is not to identify 'true' moral principles for AI; rather, it is to identify fair principles for alignment that receive reflective endorsement despite widespread variation in people's moral beliefs. The final part of the paper explores three ways in which fair principles for AI alignment could potentially be identified.

## 1 Introduction

The development and growth of artificial intelligence raises new and important questions for technologists, for humanity, and for sentient life more widely. Foremost among these is the question of what—or whose—values AI systems ought to align with. One vision of AI is broadly utilitarian. It holds that over the long run these technologies should be designed to create the greatest happiness for the largest number of people or sentient creatures. Another approach is Kantian in character. It suggests that the principles governing AI should only be those that we could rationally will to be universal law, for example, principles of fairness or beneficence. Still other approaches focus directly on the role of human direction and volition. They suggest that the major moral challenge is to align AI with human instructions,

✉ Iason Gabriel
  iason@deepmind.com

1 DeepMind, London, UK

intentions, or desires. However, this ability to understand and follow human volition might itself need to be constrained in certain ways—something that becomes clear when we think about the possibility of AI being used intentionally to harm others, or the possibility that it could be used in imprudent or self-destructive ways. To forestall these outcomes, it might be wise to design AI in a way that respects the objective interests of sentient beings or aligns with a conception of basic rights, so that there are limits on what it may permissibly do.

Behind each vision for ethically-aligned AI sits a deeper question. How are we to decide which principles or objectives to encode in AI—and who has the right to make these decisions—given that we live in a pluralistic world that is full of competing conceptions of value? Is there a way to think about AI value alignment that avoids a situation in which some people simply impose their views on others?

Before we can answer these questions, it is important to be clear about what we mean by artificial intelligence and the challenges it raises. In common language, the term 'artificial intelligence' refers both to a property or quality of computerized systems and to a set of techniques used to achieve this capability, for example, machine learning (ML). For the sake of the present discussion, 'intelligence' is understood to refer to 'an agent's ability to achieve goals in a wide range of environments' (Legg and Hutter 2007, 12).[1] The artificial nature of this intelligence can best be understood in contrast with biological intelligence: rather than being evidenced by living organisms, AI is a property attributed to computer systems and models. Artificial intelligence, then, is the design of artificial agents that perceive their environment and make decisions to maximise the chances of achieving a goal. In this context 'machine learning' refers to a family of statistical or algorithmic approaches used to train a model so that it can perform intelligent actions. When run on sufficiently powerful hardware, these techniques allow models to learn from experience, or from labelled or unlabelled data, without using explicit instructions. Innovations in ML, the collection of vast datasets, and the growth of computing power have together fostered many of the recent breakthroughs in AI research.

The goal of AI value alignment is to ensure that powerful AI is properly aligned with human values (Russell 2019, 137). Indeed, this task, of imbuing artificial agents with moral values, becomes increasingly important as computer systems operate with greater autonomy and at a speed that 'increasingly prohibits humans from evaluating whether each action is performed in a responsible or ethical manner' (Allen et al. 2005, 149). The challenge of alignment has two parts. The first part is *technical* and focuses on how to formally encode values or principles in artificial agents so that they reliably do what they ought to do. We have already seen some examples of agent misalignment in the real world, for example with chatbots that ended up promoting abusive content once they were allowed to interact freely with people online (Miller et al. 2017). Yet, there are also particular challenges that arise specifically for more powerful artificial agents. These include how to prevent 'reward-hacking', where the agent discovers ingenious ways to achieve its objective or reward, even though they differ from what was intended, and how to evaluate the performance of agents whose cognitive abilities potentially significantly exceed our own (Irving et al. 2018; Leike et al. 2018; Christiano 2016).

---

[1] This is only one definition drawn out of a diverse family of conceptions of intelligence (Legg and Hutter 2007, 11; Gardner 2011; Cave 2017).

The second part of the value alignment question is *normative*. It asks what values or principles, if any, we ought to encode in artificial agents. Here it is useful to draw a distinction between *minimalist* and *maximalist* conceptions of value alignment. The former involves tethering artificial intelligence to some plausible schema of human value and avoiding unsafe outcomes. The latter involves aligning artificial intelligence with the correct or best scheme of human values on a society-wide or global basis. While the minimalist view starts with the sound observation that optimizing exclusively for almost any metric could create bad outcomes for human beings, we may ultimately need to move beyond minimalist conceptions if we are going to produce fully aligned AI. This is because AI systems could be safe and reliable but still a long way from what is best—or from what we truly desire.

This article focuses on the normative part of the value alignment challenge. It has three parts. The first looks at the relationship between technical and non-technical aspects of AI alignment, and argues that they are not orthogonal but rather related in important ways. The second section looks at the goal of alignment in more detail. It considers whether it is best to align AI with instructions, intentions, preferences, desires, interests, or human values, and draws out salient distinctions between these aims. The third section addresses the question of alignment for groups of people that ascribe to different moral systems. I argue that the central challenge we face is not to identify the true moral theory and encode it in machines but rather to identify fair processes for determining which values to encode. This section also explores three such potentially fair processes. I conclude with an overview of the material covered in the paper and discussion of possible future research directions.

## 2 Technical and Normative Aspects of Value Alignment

What is the relationship between the technical and normative parts of the alignment challenge? Are the two tasks, of working out how to align AI with certain principles and choosing those principles, independent? Or are they related to each other in certain ways? To make progress in answering these questions, it may be helpful to consider what we might term the 'simple thesis'. According to the simple thesis, it is possible to solve the technical problem of AI alignment in such a way that we can 'load' whatever system of principles or values that we like later on. Among those who adhere to some version of this thesis, it sometimes carries with it the further tacit implication that the search for answers to the philosophical questions can be delayed. Is the simple thesis likely to be true, given what we currently know about the state of machine learning?

The discipline of ML encompasses a variety of different approaches. One branch of ML, called 'supervised learning', focuses on training a model to identify and respond to patterns using labelled data, which allows a human to evaluate the model's performance. 'Unsupervised learning', by way of contrast, aims to uncover patterns in un-labelled data and to perform tasks on that basis. However, a particularly promising approach for building more advanced forms of AI, and the approach on which I focus below, is reinforcement learning (RL). With RL, an agent learns what to do by trying to maximise a numerical reward signal that it receives from the environment. As Sutton and Barto explain, 'The agent's sole objective is to maximise the total reward it receives over the long run. The reward signal thus defines what are the good and bad events for the agent. In a biological system, we might think of rewards as analogous to the experiences of pleasure or pain' (Sutton and Barto 2017, 5). The agent then learns to maximise reward through a process of trial-and-error and refinement that, if successful, leads to better and better performance.[2]

It is important not to understate the dexterity of existing RL models. They already have a wide range of proven applications including factory robotics, commercial inventory management, and the prediction of novel protein structures from scratch (Kober et al. 2013; Evans et al. 2018). At the same time, we should recognize that these systems, taken as a whole, function as very powerful optimisers. Indeed, this tendency sits at the heart of certain long-term safety concerns about AI; if it were to optimise for something that we did not really want, then this could have serious consequences for the world (Bostrom 2016). This propensity to optimise also gives AI a certain moral valence. In general, it seems likely that it will be easier to align AI with moral theories that have the same fundamental structure based on maximizing reward over time in the face of uncertainty, than with other alternatives. Consequentialist moral theories, the most famous of which is act utilitarianism, fit the bill.

According to act utilitarianism, the morally right action to take is the one that will create the greatest happiness for the greatest number of sentient creatures in the future. In this regard, the parallels with RL are clear. As one author notes, we 'ought to immediately see that RL is very much like utilitarianism because both the RL agent and the utilitarian moral agent seek to determine which action will maximise the good, and how this dictum eventually proceeds to all agents achieving some desirable future state, goal or consequence' (Roff 2020). Indeed, some AI researchers incorporate the notion, of maximizing expected utility, into their very definition of an ideal artificial agent (Russell and Norvig 2010, 34).

By way of contrast, it is less obvious how RL can be used to align agents with non-consequentialist moral frameworks. One set of alternatives focuses not on maximizing a given value, such as happiness, but on 'satisficing'—an approach requiring only that people have *enough* of certain goods (Slote and Pettit 1984). For example, we might want AI to treat people with sufficient respect, so that it treats them well in the ways that matter, but not with excessive deference at the expense of other values. Satisficing may also represent a partial solution to safety problems associated with strong optimization. For this reason, researchers at the Machine Intelligence Research Institute have developed the idea of 'quantilizers', which represent a way of programming AI that potentially renders an agent indifferent between a top tier of good outcomes (Taylor 2016).

---

[2] To achieve this outcome, RL systems contain four core elements: a *policy*, which defines the agent's way of behaving at a given time; a *reward signal*, which defines its goal; a *value function*, which estimates the long-term sum of different states of affairs; and a *model of the environment*, which allows the agent to make predictions about how the environment will respond to its decisions (Sutton and Barto 2017, 5).

The situation is still more complicated when we come to another concept that is axiomatic for many moral theories, namely the idea of *rights* or deontological constraints. The notion that sentient creatures have rights can be understood in different ways, but, broadly speaking, rights are claims by individuals against collectives that resist aggregation and mark out a domain of things that cannot permissibly be done to them. Rights work as 'trumps' against claims about general utility (Dworkin 1984) or as 'side-constraints' on otherwise optimizing action (Nozick 1974). Although this remains to be seen, it may be difficult to robustly specify and guarantee rights-respecting behaviour on the part of agents whose learning process and decision-making are guided primarily by an optimization function.[3] Moreover, the challenge for programmers is greater still if the moral approach we ultimately endorse, and ask AI to enact, is highly complex. For example, both Kantian and contractualist moral theories require that an agent understand the concept of a 'reason' and subject it to certain kinds of hypothetical test before knowing how to proceed—capabilities that extend well beyond most existing forms of artificial agent (Kant and Schneewind 2002; Scanlon 1998).[4]

In the light of these considerations, it seems possible that the methods we use to build artificial agents may influence the kind of values or principles we are able encode. If this is the case, then the simple thesis may still identify an important design principle for AI—that we want this technology to be compatible with a wide range of perspectives and values. However, the goal of value-open design may also need to be something that the AI community consciously aspires towards and designs for.

Recognizing the difficulty of formally specifying and encoding moral principles in artificial systems, some researchers have asked whether there are technical approaches to AI alignment that could circumvent the need to directly specify moral principles for AI altogether. Part of the impetus for this line of inquiry comes from a family of ML approaches known as 'inverse reinforcement learning' (IRL). Unlike regular RL, IRL does not specify upfront the reward function that the agent aims to maximise. Instead the agent is presented with a dataset, environment, or set of examples, and focuses on 'the problem of extracting a reward function given observed optimal behaviour' (Ng and Russell 2000, 1). This can be done in a number of ways. One set of approaches focuses on *imitation* or *apprenticeship learning*,

---

[3] Perhaps the most promising approach to this challenge, which involves constrained optimization, come with the strong requirement that these constraints can be fully specified a priori and tends to respect them only 'in approximation' (Achiam et al. 2017). Independently, Arnold et al. (2017) have explored ways in which the logical descriptions of norms could potentially constrain RL systems, and Arkin et al. (2009) have suggested that AI systems include an 'ethical governor' in their model architecture which aims to ensure that final outputs are consistent with moral constraints.

[4] The capacity to reason and offer reasons for action features prominently in certain accounts of agent alignment, particularly those that focus on the integration of AI into our social world (Arnold et al. 2017). However, it would seem to require an advanced grasp of natural language and a capacity for perspective-taking or 'theory of mind' (Rabinowitz et al. 2018).

where the agent learns to infer the reward function of a human expert and then perform the task (e.g. moving a robotic arm) to a very high standard (Abbeel and Ng 2004). On another version of IRL, a model is trained to infer a reward function from agents by studying behaviour exhibited in large datasets. According to Vasquez, whose concern is with the ability of artificial agents to navigate social environments, this kind of IRL could allow AI to 'model the factors that motivate people's actions instead of the actions themselves' (Vasquez et al. 2014). A third approach to alignment involves using *evolutionary* methods (Salimans et al. 2017). These approaches evaluate the 'lifetime' behaviour of many agents, each using a different policy for interacting with its environment, and select those behaviours that are able to obtain the most overall reward (Sutton and Barto 2017, 6).

Each approach could carry over into the domain of AI value alignment in interesting ways. Apprenticeship or imitation learning could be used to learn good or virtuous conduct from a moral expert, if such a person exists and can be reliably identified—a question that forms the crux of virtue ethics (MacIntyre 2013; McDowell 1979; Vallor 2016). The second family of approaches, involving large datasets, could be used to learn values or preferences from large numbers of people and provide aggregate guidance about their preferred outcomes or beliefs concerning good conduct (Russell 2019, 177). Finally, evolutionary processes could be used to explore how different agents interact with a simulated social world, selecting and iterating upon the candidates that appear to be most moral. Each of these approaches to normative value alignment is indirect. Rather than specifying moral principles upfront, they require only that we present the agent with examples of good conduct, or that, for ethically aligned AI, 'we know it when we see it' and make decisions on that basis.

Yet, while these may be worthwhile technical projects, it should also be clear that none of these approaches avoids the need for moral evaluation altogether. Instead, the fundamental normative question of what AI ought to be aligned with simply returns in different guises. To deploy these approaches successfully we would still need to know: Who is the moral expert from which AI should learn? From what data should AI extract its conception of values, and how should this be decided? Should this data include everyone's behaviour, or should it exclude the behaviour of those who are manifestly unethical (sociopathic) or unreasonable (fundamentalists)? Finally, what criteria should be used for determining which agent is the 'most moral', and is it possible to rank entities in this way?

Thus, we encounter limits to what can be done by technologists alone. At this boundary sits a core precept of modern philosophy: the distinction between facts and values (Cohen 2003). It follows from this distinction that we cannot work out what we ought to do simply by studying what is the case, including what people actually do, or what they already believe. Simply put, in each case, people could be mistaken. Because of this, AI cannot be made ethical just by learning from people's existing choices. Of course, there may still be good reasons for the appeal of 'bottom-up' approaches to ethical alignment—trying to achieve ethically aligned AI by training it to better identify and understand human values and intuitions. However, the value alignment problem cannot be solved by inference from large bodies of human-generated data by itself.[5] Whichever technical approach we settle upon, we need greater clarity about the goal of value alignment and about appropriate moral principles for AI.

## 3 The Goal of Alignment

In the context of value alignment, the notion of 'value' can serve as a place-holder for many things. In the early days of AI research, Norbert Wiener wrote that, 'if we use, to achieve our purposes, a mechanical agency with whose operation we cannot interfere effectively… we had better be quite sure that the purpose put into the machine is the purpose which we *really desire*' (1960). More recently, the Asilomar AI Principles held that, 'Highly autonomous AI systems should be designed so that their goals and behaviours can be assured to align with *human values* throughout their operation' (Asilomar 2018). And, Leike et al. argue that the key question is 'how can we create agents that behave in accordance with the user's *intentions*?' (2018). Despite the apparent similarity of these formulations, there are significant differences between desires, values, and intentions. Which of these, if any, should AI really be aligned with?

Among the technical research community, there is a relatively clear consensus that we do not want artificial agents to follow instructions in an extremely literal way. Yet, beyond this, important questions remain. For example, do I want the agent to do what I *intend* it to do, or should it do what is *good* for me? And what should the agent do if I lack important information about my situation, engage in faulty reasoning, or attempt to implicate the agent in a process harmful to myself or others? To make headway in this area, this section aims to clarify different goals for alignment, focusing primarily on one-person-one-agent scenarios.

To begin with AI could be designed to align with:

i. Instructions: the agent does what I instruct it to do.

However, as Russell has pointed out, the tendency towards excessive literalism poses significant challenges for AI and the principal who directs it, with the story of King Midas serving as a cautionary tale (Russell 2019, 136). In this fabled scenario, the protagonist gets precisely what he asks for—that everything he touches turns to gold—not what he really wanted. Yet, avoiding such outcomes can be extremely hard in practice. In the context of a computer game called CoastRunners, an artificial agent that had been trained to maximise its score looped around and around in circles ad infinitum, achieving a high score without ever finishing the race, which is what it was really meant to do (Clark and Amodei 2016). On a larger scale, it is difficult to precisely specify a broad objective that captures everything we care about, so in practice the agent will probably optimise for some *proxy* that is not completely aligned with our goal (Cotra 2018). Even if this proxy objective is 'almost' right, its optimum could be disastrous according to our true objective.

---

[5] See, for example, Awad et al. (2018).

In the light of this, a better option could be:

ii. Expressed intentions: the agent does what I intend it to do.

Given the foregoing concern, many researchers have argued that the challenge is to ensure AI does what we really intend it to do (Leike et al. 2018). An artificial agent that understood the principal's intention in this way would be able to grasp the subtleties of language and meaning that more naive forms of AI might fail to understand. Thus, when performing mundane tasks, the agent would know not to destroy property or human life. It also would be able to make reasonable decisions about trade-offs in high-stakes situations. This is a significant challenge. To really grasp the intention behind instructions, AI may require a complete model of human language and interaction, including an understanding of the culture, institutions, and practices that allow people to understand the implied meaning of terms (Hadfield-Menell and Hadfield 2018). It therefore seems entirely correct that the research community is dedicating substantial attention to the task of closing the instruction-intention gap.

Yet it may be a mistake to think that successful value alignment ends here. To begin with, in order to align itself successfully with human intentions, advanced AI may also need to have a robust grasp of human preferences and values. This could be because of the challenge posed by the implied meaning of terms: to respond to a person's intention AI may need to understand things that are independent of the intention itself. Or it may be because the intended outcome directly references preferences or values, for example, if an agent is instructed to 'do what is best for everyone'. Furthermore, alignment with expressed intentions may prove inadequate in some situations. For example, if powerful AI systems function at super-human speed, which seems likely, then it may not be possible to provide them with immediate and continuous direction in this way (Russell et al. 2015; Soares 2014). Instead, artificial agents would need to be able to make sound decisions by default, including in unforeseen situations, without explicit instructions or well-formed intentions from a human operator.[6]

We also need to keep in mind that our intentions, even if clearly expressed, may be faulty. It is quite possible for intentions to be irrational or misinformed, or for the principal to form an intention to do harmful or unethical things. Acting on the wrong kind of intention could, therefore, lead to a variety of bad outcomes. Thus, while we may want to leave significant scope for direction through expressed intention, there may ultimately be reasons to limit the role played by intentions altogether, no matter how good an AI is at following them.[7]

---

[6] For Bratman, intentions are elements in larger plans that are typically *partial*: they have a hierarchical structure that can be filled in later as required (Bratman 1987, 50). This suggests that an AI system, operating at speed, could face choices about which explicit intentions provide only incomplete guidance.

[7] Questions about the design of more powerful AI systems are relevant here. On one vision of successful alignment, there would always be a principal that provides coherent, continuous, principled direction to the agent. The task of the agent would then be to understand and follow the instructions that the principal gives it. In this case the artificial agent functions more as a tool, performing only executive functions (Bostrom and Yudkowsky 2014, 184). On another view, continuous principled direction of this kind is

These considerations point us in the direction of a third alignment option. Perhaps AI should be designed to align with:

iii. Revealed preferences: the agent does what my behaviour reveals I prefer.

There are several versions of this view, with the most developed accounts focusing on AI alignment with preferences as they are revealed through a person's behaviour rather than through expressed opinion. In this vein, AI could be designed to observe human agents, work out what they optimise for, and then *cooperate* with them to achieve those goals (Hadfield-Menell et al. 2016).[8]

This approach has certain advantages. To begin with, it could help an artificial agent react to situations appropriately in real time. It could also help the agent navigate the social world successfully by developing sensitivity to the preferences of others, ensuring that it didn't do things that no one wanted. Moreover, a focus on revealed preferences works with data that is accessible to the agent, and the approach has been well-studied in the discipline of welfare economics.

At the same time, the proposal has several limitations. From a practical point of view, any attempt to infer a reward or utility function from observed behaviour encounters the problem of what Ng and Russell (2000) term 'degeneracy': at any moment, there is a large set of reward functions for which an observed behaviour is optimal. This means that it is very hard to make reliable inferences from observed behaviour, and it may be impossible to do so without making controversial assumptions about human rationality (Sen 1973; Armstrong and Mindermann 2018). Revealed preferences also provide reliable information only about choices that people encounter in real life. This means that it is hard to model preferences for situations that are rarely observed, even though these cases—for example, emergencies—could be morally important.[9]

On a philosophical level, the challenge is potentially greater still. It is simply not clear why we should treat revealed preferences as having weight or authority when deciding what artificial agents ought to do. The satisfaction of revealed preferences may serve as a very weak proxy for something like happiness or autonomy. However, it is also true that people make bad choices for a variety of reasons. What is good for us can, therefore, differ systematically from the preferences we reveal (Kymlicka 2002, 15).

More specifically, alignment with revealed preferences encounters the following three problems. First, people have preferences for things that *harm them*. This could happen because they do not know that their choice will have this effect, suffer from addiction, engage in severe hyperbolic discounting, or want to hurt themselves.

---

Footnote 7 (continued)

unlikely to be forthcoming. The agent would then need a more robust capability for moral decision-making.

[8] More precisely, it may be possible through observation to construct a utility curve for an individual or collective and to align the conduct of the artificial agent with this curve, which is a mathematical function that ranks alternatives according to preference. For additional challenges encountered by this approach see Eckersley (2018).

[9] Thank you to Martin Chadwick for this point.

Second, people have preferences about the *conduct of other people*. If these preferences are counted automatically, and not subjected to some further evaluative standard, then they have the potential to restrict the freedom or happiness of those people in various ways (Dworkin 1981). This can be seen in the beliefs people have about other people's sexuality or private behaviour. Moreover, some preferences are malicious: sometimes people want to harm others or to see them fail in painful ways. Third, preferences are not a reliable guide to what people really want or deserve because preferences are *adaptive*. On this point, Sen notes that 'a person who has had a life of misfortune, with very little opportunities, and rather little hope, may be more easily reconciled to deprivations than others reared in more fortunate or affluent circumstances' (1999, 45). As a consequence, they may want less because they have adapted to their situation and mistakenly believe that this is all they are entitled to hope for. By responding to preferences alone, or preferences in combination with expressed intentions, AI could therefore come to act on data that is heavily compromised and reflects entrenched discrimination. Thus, even if we could have continuous alignment with revealed preferences, there is no guarantee that this outcome would be ethical or prudent.

To address these problems, we may think it better for AI to align with:

iv. Informed preferences or desires: the agent does what I would want it to do if I were rational and informed.

By focusing on the subset of preferences that a person would have if they were both informed and instrumentally rational, AI could avoid a large number of errors that arise from limited information and poor reasoning. We might also move closer to their authentic preferences: to a set of considerations that more closely reflect what people really want or desire.

But this approach creates its own challenges. To align AI in this way, we would have to apply a corrective lens or filter to the preferences we actually observe. As a consequence, the approach is no longer strictly empiricist (see Russell and Norvig 2010, 6). Moreover, prioritizing informed preferences does little to address the challenge of self-harming or unethical preferences. According to the philosopher David Hume, instrumental rationality and full information are compatible with any type of end, including those that harm oneself or others (Blackburn 2001). Thus, even if we align AI with the preferences people would have if they were rational and informed, it may still be necessary to constrain the agent's range of permissible action in further ways.

For some philosophers and technologists, the solution to these problems resides in more substantive conceptions of human rationality and reason. According to this view—endorsed in various forms by Immanuel Kant, Derek Parfit, Amartya Sen, and others—rationality also requires us to select valid ends, of the kind that can provide a reason for action (Korsgaard et al. 1996; Parfit 2011; Sen 2004). On this view, it is not rational for a person to focus their life's energy on the task of counting the blades of grass in a field—to the detriment of relationships, physical health, and well-being—even if they form the volition to do so, because the goal is ultimately worthless (Quinn and Foot 1993). However, the conception of rationality involved here is both contested and indeterminate. Some people reject the view that rationality applies to the selection of ends, and even among those who do believe that rationality applies to that selection, there is little agreement about what ends substantive rationality tracks or requires.[10] It may be better, therefore, to focus on alignment goals that are subject to less metaphysical and practical disagreement.

One quasi-empiricist alternative would be to align AI with:

v. Interest or well-being: the agent does what is in my interest, or what is best for me, objectively speaking.

On this view, AI would be designed to promote whatever is good for a person's well-being. It would be calibrated to pursue the sorts of things that fulfil human needs, make our lives go well, and support overall flourishing. Though interest, understood in this way, is not amenable to direct scientific observation, we are able to collect data about the things that people believe constitute or contribute to their well-being, as well as information about the kinds of things that make a human life go well (Sumner 1996; Vaillant 2008). The disciplines of philosophy, psychology, and economics all contribute to our understanding of well-being in this sense.

Some philosophers have argued that well-being consists only in subjective sensory experience or in the satisfaction of an individual's desires. However, the most widespread accounts of well-being are multifaceted and include factors that can be more objectively ascertained. These factors include goods such as physical health and security, nutrition, shelter, education, autonomy, social relationships, and a sense of self-worth. Within the subfield of economics sometimes referred to as 'human development', there has been further progress mapping out this domain. Rather than assume that well-being results from the satisfaction of preferences, as classical economics somewhat dubiously maintains, the theory of human development argues that welfare stems from the ability to exercise certain core capabilities that both constitute and support human flourishing (Sen 2001). These capability-based metrics have found a measure of cross-cultural affirmation and consent (Nussbaum 1993).

So, should AI be aligned with an objective conception of human interests? This approach has much to recommend it. To begin with, although there is disagreement about the nature of well-being, the scope of this disagreement is relatively narrow. While there is fundamental disagreement about substantive conceptions of rationality, most people accept that there are core elements of human well-being, albeit with some variation across time and space. More importantly, this approach potentially addresses two of the central problems that we have encountered so far. First, AI that is aligned with underlying human interests would not be imprudent or assist in self-harm. Second, AI that is designed to respect and respond to human interests, in a group setting, is less likely to harm other people. For this reason, the idea that AI should not act in ways that are detrimental to human interests seems to represent a second rough boundary condition for its acceptable development and use, alongside safety requirements.

---

[10] For some theorists, such as Joseph Raz, the family of valid aims is quite varied (Raz 1999). For others, such as Kant and Schneewind (2002) or Smith (1994), substantive rationality converges around a set of universal ends that are largely synonymous with moral law.

Nonetheless, this picture of AI alignment is incomplete. Viewed from the perspective of a single person, the fact that something is in my interest doesn't mean I ought to do it or that I am morally entitled to do so. For instance, stealing may be in my interest, but I am not entitled to steal, except perhaps in very constrained circumstances. The same is true for collective decisions. For example, it could be wrong to use an innocent person as a scapegoat to avert violence, even if it is in the collective interest of a society to do so. By extension, it would be wrong for AI to perform these actions. These examples point to a variety of ways in which the interest-based account is insufficient. First, we need a way of deciding how to manage trade-offs between the interests and claims of different people. Unlike simple optimization, this account could factor in considerations of justice or rights. Second, we need principles for deciding whose interests or needs count for the purpose of AI alignment. Is it only the people who are currently alive, or do the interests of those not yet born also need to be factored in?[11] And is it only humans whose interests matter, or do the interests of nonhuman animals and other forms of sentient life count as well? Third, there may be other morally relevant considerations that an interest- or needs-based approach overlooks altogether, for example, the intrinsic value of the environment (Jonas 1984, 8).

These considerations prompt us to consider a final possibility. AI could be designed to align with:

vi. Values: the agent does what it morally ought to do, as defined by the individual or society.

What do 'values' signify in this context? Roughly speaking, values are natural or non-natural facts about what is good or bad, and about what kinds of things ought to be promoted. This normative sense of value differs significantly from the notion of value applied to goods or commodities in market contexts. Many things have value that is not captured by markets, or that is priced in ways that are obviously distorted (Sandel 2012). This is true for goods such as love, friendship, the environment, justice, freedom, and equality. At the same time, there is considerable metaethical debate about whether values, of a kind that are both normative and objective, actually exist (Mackie 1990). Metaethical realists maintain that they do (Nagel 1989; Parfit 2011). The alternative point of view holds that our evaluative judgments ultimately lack this factual foundation (Rorty 1993).

---

[11] Strictly speaking, alignment with human preferences or interests can generate concern for the welfare of future people *if* present generations care about them or have a stake in their welfare (Scheffler 2018). However, these considerations are likely to diminish in force when we think about more distant future people. For this reason, concern for their welfare is better anchored in the notion that they too have value and deserve to be considered when designing powerful technologies. Articulating a version of this claim, Ord writes that 'people matter equally regardless of their temporal location' (Ord 2020, 45). Jonas has similarly argued for the fundamentally moral character of our responsibility to future generations (1984, 37, 41).

The metaethical debate may seem critical. After all, if values do not have this objective basis, how can AI be developed to align with them? Yet these concerns turn out to have limited significance for the question at hand. To see why, we need to acknowledge first that, in practice, AI would have to be aligned with some set of *beliefs about value*, not with value itself. One reason it would be good to align with people's beliefs about value would be that values exist and that their beliefs reliably track or reflect this underlying reality. However, even without this assumption, it could still be best to align AI with beliefs about value, for a number of reasons. From the point of view of social psychology, values—understood as ideals shared by members of a culture about what is good or bad—play an important role in social life (Haidt 2012). They help communities of individuals resolve collective-action problems, stabilise social relationships, and flourish over time. In the words of the psychologist Greene, 'Morality is a set of psychological adaptations that allow otherwise selfish individuals to reap the benefits of cooperation' (2014, 23). It may, therefore, be prudent to align AI with a community's moral beliefs.[12] This approach would also serve to limit the prospect of alignment with malicious goals or behaviour in many cases.

A values-based approach to AI alignment has three further advantages: it can integrate different loci of alignment into a single decision-making schema, add nuance when thinking about how to evaluate aggregate claims, and include the full scope of things people actually care about. Let us take these points in turn. First, by shifting the focus of alignment to moral beliefs or principles, we avoid having to choose between alignment with expressed intentions, revealed preferences, or objective interests. Instead, it becomes possible to combine some measure of principal-direction with a set of objective constraints. For example, The Harm Principle, advocated by Mill, suggests that people should be free to act as they *wish*, unless doing so would result in *harm* to another person (1859). Asimov's Three Laws of Robotics also have this nested structure, situating the imperative to obey human orders between the absolute prohibition on harming human life and a tertiary duty of self-preservation (2004). Second, when it comes to decisions that affect groups of people, moral principles tend to replace impartial maximization with a more nuanced calculus that includes considerations such as justice and rights. Instead of simply performing whatever action maximises total well-being, we may want consider how well-off each person is relative to others, the choices they have made, and whether the proposed action violates a moral constraint. Third, a values-based approach to AI alignment can encompass considerations that both volitional and interest-based accounts overlook. For example, when hewing to principles, an artificial agent could account for the intrinsic value of the natural world, the welfare of animals, or the moral claims of people not yet been born.

In the light of these considerations, it makes sense to think that AI should be aligned with guiding principles anchored in some set of evaluative judgments (Moor 1999). Maximally aligned AI would then be AI that did what it ought to do, judged from an appropriate evaluative standpoint. However, this proposal encounters two major difficulties. The first is to specify what values or principles AI should align with. Even if we leave out religious beliefs, which would be the first port of call for much of humanity, there are many reasonable candidates to choose from. For example, AI could align with

> the aggregate interest of everyone, weighted so as to give priority to the worst-off

> or

> whatever maximally embodies the human understanding of virtue, so that the artificial agent is compassionate, generous, wise, and so on[13]

> or

> an ethic based on mutual recognition and the maintenance of collective bonds—that foregrounds the importance of human relationships.[14]

---

[12] In one interesting example, researchers used a measure of 'social value orientation' to help autonomous vehicles predict the intentions and behaviour of other drivers (Schwarting et al. 2019).

The second major difficulty concerns the individual or body of people who select the principles with which AI aligns. Given the range of possible principles, who has the right to make decisions about AI alignment and on what basis? Are we concerned with the moral beliefs of a single person, with those of a specific society, or with global moral beliefs—if such beliefs exist?

## 4 Principles for Alignment

The goal of this section is to identify principles that can govern AI such that it is aligned with human values. But before we look at the options in more detail, we need to be clear about the challenge at hand. For the task in front of us is not, as we might first think, to identify the true or correct moral theory and then implement it in machines. Rather, it is to find a way of selecting appropriate principles that is compatible with the fact that we live in a diverse world, where people hold a variety of reasonable and contrasting beliefs about value.

Taking these points in turn, it is sometimes thought that if only we could identify the true moral theory then the problem of value alignment would be solved. Moreover, some authors suggest that though we may not have succeeded in identifying such an account to date, it may be possible to do so in the future after a period of 'long reflection'—perhaps with the assistance of more powerful AI systems (Perry 2018).[15] Of course, we cannot know in advance what insight AI might enable, so it is sensible to remain agnostic about the long-term value of this technology for moral philosophy. But even if it could help us answer certain questions, it is very unlikely that any single moral theory we can now point to captures the entire truth about morality. Indeed, each of the major candidates, at least within Western philosophical traditions, has strongly counterintuitive moral implications in some known situations, or else is significantly underdetermined.

---

[13] In this vein, Vallor has argued that there are set of global 'technomoral virtues' that can help humanity successfully navigate the turbulence wrought by technological change. These virtues are understood to be culturally adaptive but also anchored in 'some enduring domain of human experience' (2016, 50, 119). Although this is not their primary purpose, they could, in principle, serve as a locus for AI alignment..

[14] This ethic would have much in common with the notion of *Ubuntu* found in philosophical traditions in sub-Saharan Africa (Metz 2007). Ubuntu philosophy builds upon the idea that 'a person is a person through other persons' and that respect for these relationships, both inter-personally and as mediated through the technologies we use, is a key component of human dignity (Mhlambi 2020; Mohamed et al. 2020). Important connections can also be draw between this principle and a feminist ethic of care (Gilligan 1993; Van Wynsberghe 2013).

Furthermore, even if this were not the case and we came to have great confidence in the truth of a single moral theory, this approach immediately encounters a second problem, namely that there would still be no way of reliably communicating this truth to others. For, as the philosopher Rawls notes, human beings hold a variety of reasonable but contrasting beliefs about value. What follows from the 'fact of reasonable pluralism' is that even if we strongly believe we have discovered the truth about morality, it remains unlikely that we could persuade other people of this truth using evidence and reason alone (Rawls 1999, 11–16). There would still be principled disagreement about how best to live. Designing AI in accordance with a single moral doctrine would, therefore, involve imposing a set of values and judgments on other people who did not agree with them.[16] For powerful technologies, this quest to encode the true morality could ultimately lead to forms of domination.[17]

To avoid a situation in which some people simply impose their values on others, we need to ask a different question:

### 4.1 In the absence of moral agreement, is there a fair way to decide what principles AI should align with?

Fortunately, there is a family of approaches situated within the field of political theory that aim to answer this question, albeit for the institution of the state rather than for AI.

These approaches start from the conviction that people are free and equal, and ask what principles people, thus situated, might reasonably agree upon. A key assumption here is that people will continue to have a variety of different values and opinions; there is no demand that they set these aside. Instead, the parties need only agree on principles to govern a specific subject matter or set of relationships.

---

[15] Among other things, advanced AI might be able to help us detect underlying coherence between different ethical systems and provide us with more information about what a world in which different principles were implemented would look like.

[16] Unfortunately, this problem is not addressed by the 'parliamentary model' of moral decision-making, which encourages individuals to assign probabilities to the likelihood that different moral theories are true and then estimate the choice-worthiness of options on that basis (Bostrom 2009; MacAskill 2016). Individuals who use this method would still need to communicate the validity of their conclusions to other people who hold a different set of reasonable credences.

[17] There is some concern that this phenomenon is already occurring within the field of AI (Ricaurte 2019; Mohamed et al. 2020).

Moreover, though they need to agree on certain principles, they may choose to endorse these principles for different religious or philosophical reasons. Their agreement therefore takes the form of an 'overlapping consensus' between different perspectives (Rawls 2001, 32). Thus, even without agreement about the fundamental nature of morality, people may still come to a principled agreement about values and standards that are appropriate for a given subject matter or domain.[18] How do these ideas apply to the identification of principles for the specific subject matter of AI?

Below are three proposals.

### 4.2 Global Public Morality and Human Rights

Starting from the premise that people are free and equal, political liberals argue that it is possible to identify principles of justice that are supported by an overlapping consensus of opinion. This tradition also accepts that non-liberal societies might endorse principles of justice based on their own internal overlapping consensus (Rawls 1999). In both situations people live under principles that they have themselves endorsed, and the problem of domination or value-imposition is largely avoided.

Adapting this idea, so that it applies to the specific subject matter of AI, could mean trying to align the artificial agent with the principles of justice endorsed by the society in which it is situated—so long as these principles command the right kind of support. This project is intuitively appealing and already underway, as societies have marshalled domestic principles of justice when evaluating socially important algorithms as biased. In the United States, for example, parole-recommendation algorithms have been criticised for departing from notions of fairness foundational to the criminal justice system (Chouldechova 2017; Koepke and Robinson 2018).

Yet efforts to align AI with domestic principles of justice can only succeed up to a point. Indeed, there are three reasons to think we may ultimately need to reach for more global principles for AI. First, there is significant variation between public conceptions of justice in different societies. Unless we embrace a naive form of cultural relativism, it cannot be the case that any such conception is permitted or that whoever happens to govern a territory acquires the right to determine how AI is used. When confronted with practices that are oppressive or harmful to human life, it may be necessary to build in constraints on new technologies, even if these practices command widespread support within a given society. Second, even if we restrict the scope of alignment only to domestic principles of justice that do not serve harmful or oppressive ends, this does not guarantee congruence at the global level. In fact, ordinary moral disagreement between groups has been a persistent source of conflict in human life (Haidt 2012). On this point, Greene notes that decent, morally motivated people often come into conflict precisely *because* they hold different commonsense views about what is right, a phenomenon he terms 'the tragedy of common sense morality' (Greene 2014). Broader forms of agreement are needed to avoid recreating this outcome (Vallor 2016, 9). Third, advanced AI may well be a global technology, the operation of which cannot easily be disaggregated or packaged in the way that the domestic alignment approach supposes. The Internet provides an illuminating example. While different governments and providers have intervened to influence its character and use in different parts of the world, the technology embodies certain design principles and protocols that remain consequential despite these efforts at localisation (Leiner et al. 2009, 12). Bearing these considerations in mind, it would be helpful to develop and design AI that aligns with principles supported by a global overlapping consensus of opinion. Do principles of this kind exist?

---

[18] Rawls writes, 'In a constitutional democracy the public conception of justice should be, as far as possible, independent of controversial philosophical and religious doctrines… the public conception of justice is to be political, not metaphysical' (1985, 223).

For many theorists, the answer is yes, and the principles in question can be found in the doctrine of universal human rights (Cohen 2010; Donnelly 2007; Ignatieff 2001). Despite the existence of value pluralism, these theorists suggest that there are certain things that most people agree upon in practice, including the notion that individuals deserve some measure of protection from physical violence and bodily interference, regardless of the society they happen to live in. This common ground may also include the idea that people are entitled to certain basic goods such as nutrition, shelter, health care, and education. Indeed, the notion that people have a right to these goods has been codified in international law. Furthermore, as both a legal and philosophical doctrine, human rights have been endorsed by different groups for different reasons. Some people favour universal human rights because they believe that human life is sacred, while others see human rights as products of a contract between state and citizen, and still others see in human rights a tried and tested way of promoting welfare and minimizing harm. Lastly, the idea of human rights has significant cross-cultural support, with justifications found in African, Islamic, Western, and Confucian traditions of thought (Cohen 2010, 335–343). Thanks to this convergence in moral reasoning across cultures, Donnelly concludes that 'human rights can, and in the contemporary world do, have multiple and diverse "foundations"' (2007, 292).

The idea of human rights-congruent AI therefore has much to recommend it. If there is a global overlapping consensus concerning human rights, then AI can be aligned with human rights doctrine while avoiding the problems of domination and value imposition. Furthermore, international human rights also have a fairly good track record in practice. Efforts to codify and enforce human rights have had some success curbing state violence and other threats to human life around the globe (Risse-Kappen et al. 1999). It is possible that they could prove similarly effective in curbing the risks associated with digital technologies (UN Secretary General 2019).

Nonetheless, even if we agree that human rights have an important role to play in value alignment, a number of questions still need to be addressed. To begin with, *which* human rights should AI be aligned with? Are we concerned only with negative duties not to harm people or also with positive obligations to ensure that they can access vital goods and services? In this regard we appear to face a trade-off. On the one hand, negative rights are widely endorsed but have limited scope. They rule out a certain class of actions but do not provide guidance in all situations, for example, when determining what goals to prioritise. On the other hand, positive rights address this limitation, providing designers with a richer set of goals and aspirations, but command significantly less global support in practice. Another challenge stems from the fact that the doctrine of human rights applies primarily to the specific political relationship between states and individuals, so it is unclear how it translates directly into guidance for artificial agents. Indeed, the philosopher Shue has argued that the purpose of human rights is to provide human beings with a social guarantee against certain 'standard threats' and notes that kinds of threat that meet this threshold may change over time (1996). If digital technologies lead to the emergence of new standard threats, then people may be owed forms of protection that differ from those they have historically claimed against the state. The precise character of AI alignment with human rights, therefore, needs to be mapped out more fully.

Another approach to the development of an overlapping consensus in the domain of AI ethics focuses instead on existing proposals, and looks for meaningful convergence between the various codes that have been put forward. Indeed, the past five years have seen a truly vast array of proposals emerge in this domain, with Jobin et al. (2019) conducting content-analysis of 84 different documents. Among other things, these authors found evidence of a 'global convergence' around five ethical principles for AI which were: transparency, justice and fairness, non-maleficence, responsibility, and privacy. Relatedly, the multilateral Organization for Economic Cooperation and Development (OECD) has endorsed four principles as the basis for ethical AI—respect for human autonomy, prevention of harm, fairness, and explainability—building upon a similar process of consultation by Floridi et al. (2018).[19] Although these two initiatives are not identical in terms of the final output, the convergence they display is promising from the vantage point of emergent norms for AI alignment. However, in order to properly understand the significance of these developments two further questions need to be answered. First, is there genuine agreement about the meaning of the key terms and values employed here, or do they function as 'placeholders' that conceal potentially significant forms of disagreement? Second, is this convergence of opinion truly global, or is it geographically and culturally parochial in terms of who is represented?

Taking these questions in turn, Jobin et al. (2019) observe that beneath the surface there continues to be 'substantive divergence in relation to how these principles are interpreted, why they are deemed important, what issues, domains or actors they pertain to, and how they should be implemented' (389). From the vantage point of an overlapping consensus, not all of these elements are problematic. Indeed, it is a good thing if different parties agree on shared principles for AI for different reasons. However, disagreement about what the principles mean, to whom they apply, and about how they should be implemented, raises challenges. By way of illustration, people may agree that the concept 'justice' or 'fairness' is important for AI, but disagree fundamentally about whether it applies only to processes—for example, the idea AI systems should not suffer from harmful bias—or also to outcomes, including who benefits and who loses out from this new technology (Kalluri 2020). These options point to very different requirements for aligned systems. Commenting on this difficulty, Mittlestadt (2019a, 5) notes that existing codes largely contain 'abstract and vague concepts, for example commitments to ensure AI is 'fair', or respects 'human dignity', or enables 'human flourishing', which are not specific enough to be action-guiding'. As a consequence, he argues that they do 'not reflect a meaningful consensus on a common practical direction for 'good' AI development', and that 'we must therefore hesitate to celebrate consensus around high-level principles that hide deep political and normative disagreement' (2019a, 9).

---

[19] The approach adopted by these authors has partial foundations in 'principlism' which is an approach commonly found in medical ethics (Beauchamp and Childress 2001; Mittelstadt 2019b).

With regard to the breadth and inclusiveness of existing proposals, concerns have also been raised. To begin with, out of all the codes surveyed by Jobin et al. (2019), the largest proportion came from corporations and private industry. This represents a challenge to the ideal of an overlapping consensus if the underlying goal is to identify potential convergence between the views of people who are affected by this technology. Moreover, these codes have tended to have a pronounced geographical focus, arising from actors who are located in economically developed countries. On this point Jobin et al. (2019) found that 'the underrepresentation of geographic areas such as Africa, South and Central America and Central Asia indicates that global regions are not participating equally in the AI ethics debate, which reveals a power imbalance in international discourse' (396).

Taken together, these concerns suggest that talk about an overlapping consensus of opinion, in the space of AI alignment, may be premature. In order for it to emerge, efforts to derive principles for aligned AI need to be genuinely intercultural and inclusive, offer concrete guidance across a range of situations, and be stable over time, meaning that people continue to affirm and endorse these principles once they have seen them implemented in practice (Rawls 1987, 11).

### 4.3 Hypothetical Agreement and the Veil of Ignorance

A second approach to pluralistic value alignment focuses not on the values people already agree on, but rather on the principles they would agree upon if they were placed in a position where no one could impose their view on anyone else. To understand what principles would be chosen in this kind of situation, Rawls proposes a thought experiment in which parties select principles from behind a 'veil of ignorance'—a device that prevents them from knowing their own particular moral beliefs or the position they will occupy in society. Behind the veil, 'the parties… do not know how the various alternatives will affect their own particular case and they are obliged to evaluate principles solely on the basis of general considerations' (Rawls 1971, 137). The outcome of deliberation under these conditions is principles that do not unduly favour some over others. Such principles are therefore, *ex hypothesi*, fair. Adapting this methodology to the present case we can ask what principles would be chosen to regulate AI. What principles would people choose to regulate the technology if they did not know who they were or what belief system they ascribed to?

To answer this question, we need a clearer picture of the technology that the contracting parties are to choose principles for. Such clarity is hard to come by because there are many different kinds of AI and visions about how the technology might evolve in the future. Important archetypes include AI as a personal assistant to individuals, AI as a technology owned and operated by corporations as a consumer service, and AI that takes on an increasingly important public function via its integration with education, healthcare, and welfare systems. Moreover, there is disagreement about the level of autonomy AI systems may come to embody. Whereas the view of AI as a personal assistant tends to assume that it will remain a 'tool' which lacks real agency, RL agents take a more creative approach to problem-solving and, in this regard, embody more 'degrees of freedom' than simple rule-based systems (Dennett 2003, 162). These distinctions are important because, just as we would choose different principles to govern the behaviour of individuals, corporations, states, and supranational entities, so too would we choose different principles to govern the behaviour of different forms of AI.

Clearly it is too soon to say which, if any, of these forms of advanced AI will be created. Moreover, it seems likely that different governing principles would be chosen from behind the veil of ignorance in each case. Indeed, a virtue of this approach is its sensitivity to variation of this kind (Cohen and Sabel 2006). At the same time, there are certain principles that people situated behind the veil of ignorance might endorse across a wide range of technologies and outcomes (Moor 1999). To begin with, they would probably agree that AI should be designed in a way that is safe, reducing the risk of accident and misuse. It also seems likely that they would want to rule out a class of actions that nobody benefits from. Somewhat more tentatively, they might affirm the importance of opportunities for human control, not only as a component of AI safety but also because it reflects the value of their own autonomy or freedom. Regardless of where they are situated in relation to this technology, they would still know that they were people with their own aspirations and lives to lead. Finally, there may be certain distributive principles that would be chosen to regulate advanced AI. Without knowledge of their wealth or social standing, decision-makers might oppose large gaps between AI's beneficiaries and those who lose out from the technology. These concerns would move them in the direction of egalitarian or prioritarian principles of justice, a strong version of which would be to insist that AI must work to ensure the greatest benefit to the least well off. To meet this condition in a global context, AI would need to benefit the world's worst-off people before it could be said to be value-aligned.

### 4.4 Social Choice Theory

A third approach to pluralistic value alignment aims not to find principles we all agree on but instead to add up individual views fairly. The proposed mechanisms for doing so draw largely from social choice theory, a body of research that focuses on how to aggregate information from individuals to make collective judgments. Social choice theory has purchase both in welfare economics, where the concern is often to satisfy the preferences of a majority, and in the context of voting, where there are different ways to make collective decisions. In the context of AI, Prasad has argued that, 'Given there is no universal agreement, even among humans on ethical values, social choice is a necessary tool to address the value alignment problem' (2018, 1). Moreover, the 'bottom-up' approach to AI ethics, considered earlier, may also require determinations of this kind (Allen et al. 2005; Wallach and Allen 2008).[20]

According to 'aggregationist' approaches to value alignment, aligned agents should be designed so as to respond appropriately to ethical preferences or volitions in real time. To do this, artificial agents would need to collect the relevant data and combine it in a way that produces decisions aligned with what people really value or desire. In terms of the mechanism by which preferences or desires are summed or combined, it may be possible to borrow insights from welfare economics—given its focus on the relationship between individual and collective well-being. However, the obstacles this approach encounters are formidable. As we have seen, efforts to maximise preference satisfaction face inherent limitations: preferences are an unreliable guide to what people want, and maximization is blind to considerations of distributive justice. Some of these shortcomings may be addressed by introducing equity principles into the moral calculus, or by focusing on a subset of 'ethical preferences' or informed desires (Armstrong 2019). However, even with these measures in place, the aggregationist approach encounters further difficulties.

First, individual preferences are often inconsistent, violating basic axioms of rationality such as transitivity. This interferes with efforts to order them systematically, a constraint that also applies to people's desires or ethical beliefs, given that they too display inconsistency and variation depending upon how a choice is described (Kahneman and Tversky 2000, 168–69). Second, starting with Condorcet and building on pioneering work by Kenneth Arrow, social choice theory has identified a large number of 'impossibility theorems', which show that any rules for consistently ranking states of affairs on the basis of individual orderings will violate certain 'very mild conditions of reasonableness' (Sen 2018, 4). Similar problems have been modelled for ethical choices that involve various levels of welfare and population size (Arrhenius 2000). Social choice processes therefore seem unable to settle deeper questions about standing, measurement, and aggregation (Baum 2017, 4).

The 'democratic' approach to value alignment takes a different path. Instead of aiming for the continuous aggregation of different views, it aims to arrive at principles or goals for value alignment through voting, discussion, and civic engagement. Prasad envisages something like this, developing the idea that AI embodies tiers of decision-making authority (2018). While we may choose to delegate authority when deriving rules that help AI implement low-level goals or objectives, the higher-level rules or 'constitution' of AI—which determine the agent's fundamental goals, behaviour and internal governance—need stronger forms of endorsement.[21] Voting is a popular form of endorsement, for a number of reasons. From an epistemic point of view, it is often thought to lead to better group decisions (Estlund 2009). Majority voting is also often thought to reflect the value of autonomy because more people tend to end up living under rules they have themselves endorsed (Waldron 1999), and the idea of one-person-one-vote can be understood to embody the value of equality. Moreover, democratic processes have the potential to confer *legitimacy* on decisions about AI alignment; they can move us beyond the notion that certain principles are justified, and show, additionally, that they have been actively endorsed. This makes the principles binding in a way they would not otherwise be the case (Simmons 1999). While legitimacy may not be a concern for less powerful forms of AI, technologies that operate and make decisions at a national or international scale may need this kind of endorsement.

---

[20] The same can be said for Yudkowsky's notion that AI should be designed to align with our 'coherent extrapolated volition' (CEV). CEV represents an integrated version of what we would want 'if we knew more, thought faster, were more the people we wished we were, and had grown up farther together' (2004, 6).

[21] Parallels can be drawn between this view of AI and the structure of a corporation. Although as a body of theory, corporate governance has been less thoroughly studied than for the institution of the state, there may be lessons from this domain that carry over to questions about the internal governance of AI.

The notion that people could vote to endorse principles for AI alignment leads to a number of theoretical and practical questions. What is the best process for voting? In what forum could these decisions be made? Who should be represented—is it individuals or states—and how? At this point, we may encounter some of the paradoxes of social choice theory again, raising the prospect of regress. However, a democratic discourse could potentially be used to evaluate the strengths and weaknesses of different proposals, garnering support for those that perform best. As Sen notes, it is unlikely that there will ever be a perfect system for scaling individual interests, desires, or values to the collective level. Rather, the process of crafting rules and protocols is an art. When it comes to democratic design, we must all work to overcome these constraints and figure out what an appropriate way to align AI with moral principles looks like.

## 5 Conclusion

This paper has advanced several claims. To begin with, I argued that the machine learning techniques we use for alignment, and the values we align with, are not fully independent of one another. The way we build AI is likely to influence the values we are able to load, and a clearer understanding of the value dimension can shape AI research in productive ways. A further consequence of this is that there is no way to 'bracket out' normative questions altogether. Instead they should form part of a combined research agenda. Turning then to the goal of alignment, I argued that we should not aim to align AI with instructions, expressed intentions, or revealed preferences alone. Properly aligned AI will need to take account of different forms of unethical or imprudent behaviour, and incorporate design principles that prevent these outcomes. One way to do this would be to build in objective constraints on what artificial agents may do. More useful still, would be a set of principles that situate human direction within a moral framework that is widely endorsed despite the existence of different belief systems. This requires work both in terms of the technical specification of concepts from which principles are assembled, and also to identify principles of the right kind. The third section focused on how such principles could be selected and justified. In this regard I argued that the major challenge for normative value alignment is not to identify the true moral theory and then program it in machines, but rather to identify principles for AI that are widely held to be fair.

The problem of alignment is, in this sense, political not metaphysical. To address it, I recommended that we look more closely at principles that would be supported by a global overlapping consensus of opinion, chosen behind a veil of ignorance and/or affirmed through democratic processes.

Each of these approaches can potentially be combined in certain ways. In addition to identifying the content of principles for alignment, we should also think about how this process of integration and consensus-seeking could be carried out. Ideally, the process used to identify principles for AI alignment would be procedurally fair, in the sense of not conferring arbitrary advantage upon one party; concrete, in the sense that it produces detailed guidance; stable and robust, leading to principles that can be sustained over time; comprehensive, leaving few gaps in coverage; and genuinely inclusive, incorporating the reasoned opinions of all who are willing to cooperate in this venture. Another important quality of the process would be its ability to deal with the possibility of widespread moral error. History provides us with many examples of serious injustice that were considered acceptable by the people at the time. Given that we too may be making errors of this kind, it would be a mistake to tether AI too closely to the morality of the present moment. Finally, the paper has treated AI as an emerging technology. However, the development of different forms of AI is not inevitable. Technologists therefore face important choices about what they want to build and why. Given the potential for AI to profoundly affect our world, these too are salient questions for our time.

**Acknowledgements** This paper has benefited greatly from the input and advice of a large number of people. Within DeepMind, particular thanks are owed to Jan Leike, Laurent Orseau, Victoria Krakovna, Ramana Kumar, Tom Everitt, Marcus Hutter, Richard Ngo, Pedro Ortega, Martin Chadwick, Neil Rabinowitz, Silvia Chiappa, Tom Schaul, Andrew Trask, Sean Legassick, William Isaac, Laura Weidinger, Courtney Biles, Vishal Maini, Dorothy Chou, Koray Kavukcuoglu, and Shane Legg. Outside of DeepMind, I owe a significant debt of gratitude to Nick Bostrom, Toby Ord, Ed Felten, James Manyika, Diane Coyle, Gillian Hadfield, Joshua Cohen, Heather Roff, Shannon Vallor, Henry Shue, Joshua Greene, Rob Reich, Seth Lazar, Miles Brundage, Michiel Bakker, Vafa Ghazavi and Marie-Therese Png. Lastly, I would like to thank audiences at the Stanford Centre on Philanthropy and Civil Society (PACS), Princeton Center for Information Technology Policy (CITP), Warwick Centre for Ethics, Law and Public Affairs (CELPA), Berkeley Center for Human-Compatible AI (CHAI), and the Partnership on AI (PAI), for helpful feedback received during workshops and seminars held at these institutions.

**Funding** The research was not funded by any independent grant or award.

**Compliance with Ethical Standards**

**Conflicts of Interest** There are no conflicts of interest to declare. All contributions are my own.